\documentclass[12pt]{article}
\usepackage{amsfonts,amsthm,amsmath,amssymb,upgreek,bm}
\usepackage{hyperref}
\usepackage[paper=letterpaper,margin=.85in]{geometry}
\usepackage{graphicx}
\usepackage{units}
\usepackage{float}
\usepackage[export]{adjustbox}
\usepackage{xcolor}
\usepackage[shortlabels]{enumitem}

\newcommand{\be}{\begin{equation}}
\newcommand{\ee}{\end{equation}}
\renewcommand{\tilde}{\widetilde}
\renewcommand{\i}{\mathrm{i}}
\renewcommand{\d}{\mathrm{d}}

\numberwithin{equation}{section}

\begin{document}

\thispagestyle{empty}

 \vspace*{2.5cm}
 \begin{center}

{\bf {\LARGE A Mirzakhani recursion\\ \vspace{10pt} for non-orientable surfaces}}\\

 \begin{center}

 \vspace{1cm}

 {\bf Douglas Stanford}\\
  \bigskip \rm

 \bigskip 

 Stanford Institute for Theoretical Physics,\\Stanford University, Stanford, CA 94305

 \rm
   \end{center}

 \vspace{2.5cm}
 {\bf Abstract}
 \end{center}
 \begin{quotation}
 \noindent

We review Mirzakhani's recursion for the volumes of moduli spaces of orientable surfaces, using a perspective that generalizes to non-orientable surfaces. The non-orientable version leads to divergences when the recursion is iterated, from regions in moduli space with small crosscaps. However, the integral kernels of the recursion are well-defined and they map precisely onto the loop equations for a matrix integral with orthogonal symmetry class and classical density of eigenvalues proportional to $\sinh(2\pi\sqrt{E})$ for $E>0$. The recursion can be used to compute regularized volumes with a cutoff on the minimal size of a crosscap.
 \end{quotation}

 \setcounter{page}{0}
 \setcounter{tocdepth}{2}
 \setcounter{footnote}{0}
 \newpage

 \parskip 0.1in
 
 \setcounter{page}{2}
 \tableofcontents


\section{Introduction}
The moduli spaces of bordered orientable Riemann surfaces have finite volume, and these volumes can be computed efficiently using a recursion relation discovered by Mirzakhani \cite{mirzakhani2007simple}.

For non-orientable Riemann surfaces, the correspoding moduli spaces have infinite volume, due to a divergence associated to small crosscaps \cite{Norbury,Gendulphe}. A ``crosscap'' is a basic building-block of non-orientable hyperbolic surfaces. It consists of a geodesic hole in the surface that has been closed off by imposing an antipodal identification, as shown at left here:
\be
\includegraphics[valign = c, scale = 1.25]{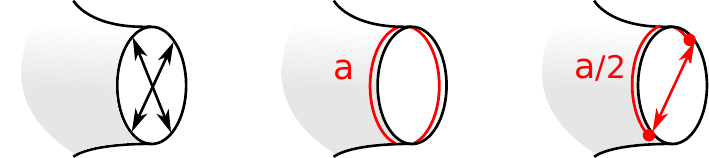}
\ee
This identification leads to a smooth but non-orientable space. In hyperbolic geometry, the crosscap can be characterized by the length, $a$, of a geodesic that winds once around the hole, or by the length $a/2$ of a geodesic that winds one-half of the way around the hole, and is closed by virtue of the identification.\footnote{The geodesic of length $a/2$ is called ``one-sided'' (as opposed to two-sided), because a neighborhood of it is topologically a Mobius strip, instead of a cylinder.}

In the moduli space integral the measure for the size of a crosscap, $a$, is \cite{Norbury,Gendulphe,Stanford:2019vob}\footnote{Assuming the standard normalization $\d b\wedge \d\tau$ of the Weil-Petersson form.}
\be
\frac{\d a}{2\tanh\frac{a}{4}},
\ee
and the integral diverges near $a = 0$.\footnote{This kind of argument is sometimes subtle: even in ordinary moduli space integrals, there is an apparent divergence at large length (or twist), which is tamed by a restriction of the moduli to a fundamental domain. Unfortunately, this doesn't help with divergences associated to {\it short} lengths. For example, in the present case we can choose the length of the shortest one-sided geodesic $a/2$ on the surface as one of the moduli. The $a\to 0$ region will then definitely be part of the integration domain.} In this paper we will show that, despite this divergence, a non-orientable version of Mirzakhani's recursion exists and the integral kernels that appear are finite. The crosscap divergence enters when we try to actually carry out the integrals to iterate the recursion. 

Still, what use is a recursion if it diverges when you try to iterate it? One answer is that the divergence can be avoided if we instead compute the volumes of an $\epsilon$-regularized version of moduli space, where one-sided geodesics are required to have length at least $\epsilon$ \cite{Gendulphe}. For small $\epsilon$, the same recursion works, with a modification to the initial conditions.

Another answer is that this recursion confirms a conjecture in \cite{Stanford:2019vob} that the non-orientable recursion should map onto the loop equations of a double-scaled matrix integral with spectral curve $y(z)^2 \propto \sin^2(2\pi z)$ and with orthogonal symmetry class. Mirzakhani's original recursion was previously shown to map onto the loop equations of a double-scaled matrix integral with the same spectral curve but unitary symmetry class \cite{eynard2007weil,Saad:2019lba} -- a fact that underlies the duality of that matrix integral with Jackiw-Teitelboim (JT) gravity \cite{Teitelboim:1983ux,Jackiw:1984je}. Formally, our result means that an unoriented version of JT gravity is also dual to a closely related matrix integral, as suggested in \cite{Stanford:2019vob}. This duality is formal because both sides are affected by the crosscap divergence. In the matrix context, the $\epsilon$-regularization can be implemented by replacing the spectral curve with the $(2,p)$ matrix integral with a large value of $p \sim 1/\epsilon$. 

\section{The volume of moduli space}
The moduli space of an {\it orientable} surface has a symplectic structure and an associated measure known as the Weil-Petersson measure. An important property of this measure is that it is invariant under the action of the mapping class group (MCG).

The moduli space of a {\it non-orientable} surface does not have a symplectic structure, but Norbury proposed a generalization of the Weil-Petersson measure for this case \cite{Norbury} and showed that it was MCG-invariant. Gendulphe later showed that Norbury's measure is actually the only invariant measure that has a factorized form in terms of length-twist variables \cite{Gendulphe}.

Another motivation for this measure comes from thinking about the path integral of Jackiw-Teitelboim (JT) gravity. In principle, this is an integral over all surfaces of a given topology, but JT gravity localizes to the moduli space of constant-curvature hyperbolic surfaces. The induced integration measure on this moduli space is a ratio of determinants known as the torsion. This was shown by explicit calculation \cite{Stanford:2019vob} to agree with Norbury's measure.\footnote{Agreement also follows more abstractly from \cite{Gendulphe} and the fact that the torsion has a factorized form in length-twist coordinates (due to the gluing rule for torsion). Thanks to Edward Witten for clarifying this.}

Norbury's measure is defined as follows. Take a possibly non-orientable hyperbolic surface with $n$ boundaries, and cut it into three-holed spheres, where some of the holes may be closed off with crosscaps, and some of the holes may be glued together with orientation-reversing reflection operators (shown as dotted lines):
\be\label{gluing}
\includegraphics[valign = c, scale = 1.5]{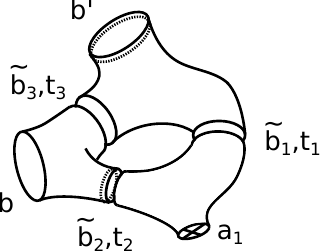}
\ee
In the above example, we have $n = 2$ boundaries and genus $g = 3/2$.\footnote{Here we are using the convention that crosscaps contribute one-half and handles contribute one to the genus. This is equivalent to defining the genus in terms of the Euler characteristic to maintain $\chi = 2-n-2g$.} The moduli space is described by the parameters in the gluing, namely the lengths $\tilde{b}_i$ and twists $\tau_i$, along with the lengths of the crosscaps $a_i$.\footnote{We use the tilde to differentiate the gluing parameters $\tilde{b}$ from the fixed lengths of boundaries $b$.} Norbury's measure is
\be
\d \tilde{b}_1 \d\tau_1 \dots \d \tilde{b}_k \d\tau_k \frac{\d a_1}{2\tanh(\frac{a_1}{4})}\dots \frac{\d a_\ell}{2\tanh\frac{a_\ell}{4}}.
\ee
The volume of the moduli space is given by integrating this measure over the possible values of the gluing parameters, up to the identifications due to the mapping class group. Informally, this means that we avoid overcounting due to the fact that it is possible to cut open the exact same surface in different ways.\footnote{It is important to emphasize that in practice, implementing the quotient due to the mapping class group is impractically difficult, except in a few very simple cases. This is why Mirzakhani's recursion is so valuable.} In addition to this integral, we sum over possible insertions of reflection operators at each gluing locus and at each external boundary, dividing by two to account for the ambiguity in the overall choice of orientation. This is the correct prescription for JT gravity with reflection symmetry gauged. 

Note that the total number of integration variables $\tilde{b},\tau$ and $a$ is invariant under the mapping class group, and determined by the genus and the number of boundaries,
\be
2k + \ell = 6g+2n-6,
\ee
but different gluing schemes may have a number of crosscaps that differs by an even integer (and a correspondingly different number of $\tilde{b},\tau$ pairs). The simplest example of this is at $(g,n) = (1,1)$ where the non-orientable contribution (a Klein bottle with a hole in it) can be constructed from a three-holed sphere either by closing off two of the holes with crosscaps, or by gluing two of them together with an antipodal reflection operator inserted.

\section{Mirzakhani's recursion and a generalization}

\subsection{Review of Mirzakhani's recursion in the orientable case}
Mirzakhani's recursion computes the volumes $V_g(b_1,\dots,b_n)$ of the moduli space at genus $g$ with $n$ external geodesic boundaries with fixed lengths $b_1,\dots,b_n$. Here we discuss one derivation of the recursion, which will generalize easily to the non-orientable case. 

It is convenient to start out by thinking not about $V_g(b_1,\dots,b_n)$, but instead about this volume times the length of one of the boundaries:
\begin{align}\label{bV}
b\cdot V_{g}(b,B) &=  \int_0^b\d p\int \frac{\d(\text{moduli of $X$})}{\text{MCG}(X)}.
\end{align}
On the LHS, $b = b_1$ is the length of an external boundary that we are singling out, and $B = \{b_2,\dots,b_n\}$ represents the lengths of the other fixed external boundaries. On the RHS we are integrating over the position of point $p$ on the first boundary and over the moduli of the surface $X$, quotiented by the mapping class group.

The key idea in Mirzakhani's approach is to consider the geodesic $\gamma_p$ that starts out at point $p$, initially moving orthogonally to the boundary. Crucially, $\gamma_p$ determines a particular three-holed sphere that can be used in the gluing construction of $X$. To see this, follow $\gamma_p$ until it either returns to the original boundary, self-intersects, or hits another external boundary; terminate $\gamma_p$ at this point.\footnote{It is also possible that none of these happen and the geodesic spirals asymptotically towards some closed geodesic; this possibility can be ignored because the set of boundary points where it happens has measure zero.} A neighborhood of this $\gamma_p$, together with a neighborhood of the boundary that it starts on (and also of the boundary that it ends on, if applicable), forms a thin ``ribbon'' shape with the topology of a three-holed sphere. Minimizing the lengths of the holes within their homology classes gives a proper three-holed sphere with geodesic boundaries, $\Lambda$. One of the boundaries of $\Lambda$ will be the original boundary of length $b$. The other two boundaries might both be ``gluing boundaries,'' or one might be external and one might be gluing.

If we didn't have $\gamma_p$, this three-holed sphere $\Lambda$ would just be one of infinitely many that we could have used as part of a gluing construction of $X$.  But using the geodesic, a particular $\Lambda$ is singled out by each boundary point $p$. In other words, the point $p$ prefers a particular decomposition of the surface $X$ into $\Lambda$ and $Y = X\setminus\Lambda$. Mirzakhani's recursion is the corresponding decomposition of the moduli space integral.

To write the exact equation we get, it is convenient to introduce the notation $F$ to describe the set of possible topologically distinct fates of the geodesic $\gamma_p$. In the orientable case, the fates we need to consider are listed below, together with example sketches:
\begin{enumerate}
\item $\gamma_p$ makes it to another external boundary
\be
\includegraphics[valign = c, scale = 1]{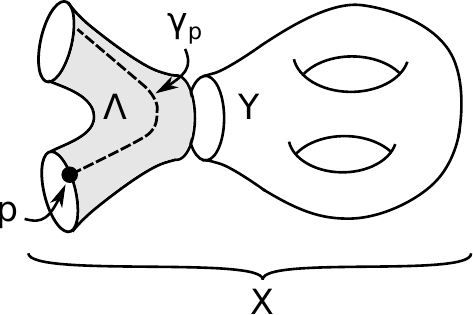}
\ee
\item $\gamma_p$ self-intersects, $\Lambda$ has two external boundaries and one gluing boundary
\be
\includegraphics[valign = c, scale = 1]{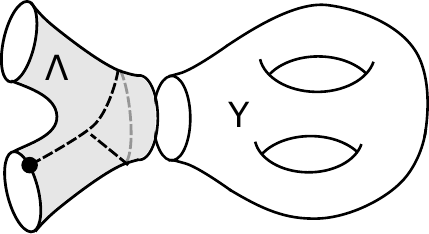}
\ee
\item $\gamma_p$ returns to the original boundary, $\Lambda$ has two external boundaries and one gluing boundary
\be
\includegraphics[valign = c, scale = 1]{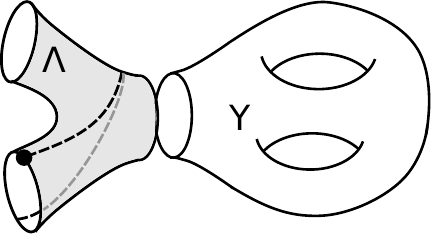}
\ee
\item $\gamma_p$ self-intersects, $\Lambda$ has one external boundary and two gluing boundaries. $Y$ might consist of one or two components:
\be
\includegraphics[valign = c, scale = 1]{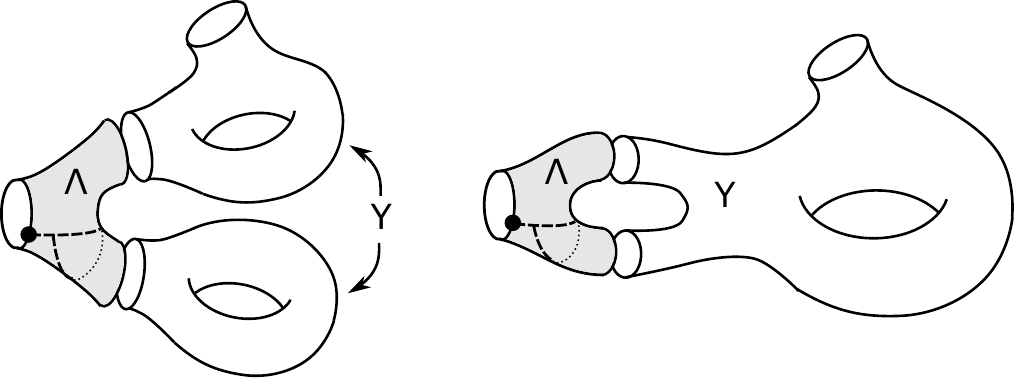}
\ee
\item $\gamma_p$ returns to the original boundary, $\Lambda$ has one external boundary and two gluing boundaries. $Y$ might consist of one or two components:
\be
\includegraphics[valign = c, scale = 1]{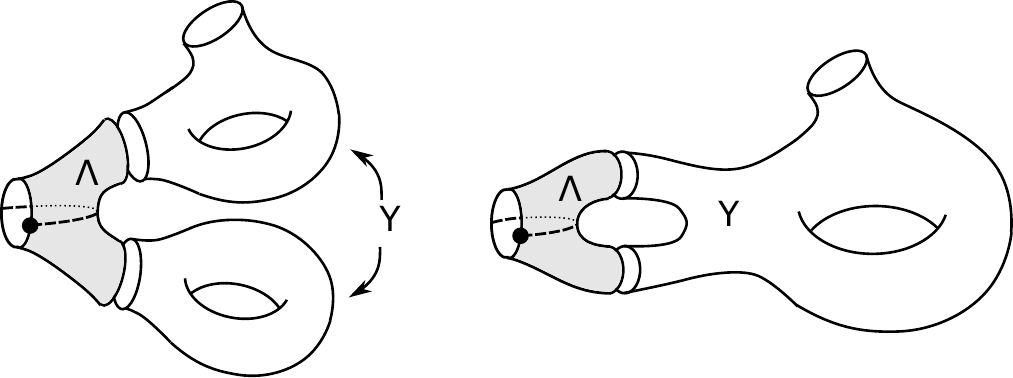}
\ee
\end{enumerate}

We refer to these five possibilities as the set of fates $F$, and the integral over the moduli space and point $p$ decomposes into a sum over $f \in F$:
\begin{align}\label{bV2}
bV_g(b,B)&=\int_0^b\d p\sum_{f\in F} \int\d\left(\text{\parbox{6.2em}{\centering\linespread{.5}\selectfont moduli gluing $\Lambda$ to $Y$}}\right) \theta\left(\text{\parbox{5em}{\centering\linespread{.5}\selectfont $\gamma_p$ has fate $f$ within $\Lambda$}}\right) \int \frac{\d (\text{moduli of $Y$})}{\text{MCG}(Y)}.
\end{align}
Crucially, this only depends on the mapping class group of $Y$, not of the full space $X$. This is because $\Lambda = X\setminus Y$ is uniquely singled out by point $p$ via the geodesic $\gamma_p$. The only restriction on the gluing parameters of $\Lambda$ comes from requiring that if we follow $\gamma_p$ within $\Lambda$, it {\it actually has fate $f$}. This is implemented by the $\theta$ factor, which is notation for an indicator function that is one if its argument is true, and zero otherwise. With this factor inserted, we can integrate freely over the parameters associated to $\Lambda$, taking into account only the obvious perodicity in the twist.

It is convenient to change the order of integration:
\begin{align}\label{bV3}
bV_g(b,B)&= \sum_{f\in F} \int\d\left(\text{\parbox{6.2em}{\centering\linespread{.5}\selectfont moduli gluing $\Lambda$ to $Y$}}\right) \int_0^b\d p\,\theta\left(\text{\parbox{5em}{\centering\linespread{.5}\selectfont $\gamma_p$ has fate $f$ within $\Lambda$}}\right) \int \frac{\d (\text{moduli of $Y$})}{\text{MCG}(Y)}.
\end{align}
Now the $\theta$ function restricts the integration range over $p$, rather than the integral over the gluing moduli of $\Lambda$. To see how this works in detail, let's consider fate 1, where $\gamma_p$ travels from the original boundary to one of the other external boundaries, say the one of length $b_k$. Then $\Lambda$ is a three-holed sphere with boundaries of length $b,b_k$ and $b'$, and the gluing moduli consist of $b'$ and the corresponding twist $\tau'$. The contribution of fate 1 to (\ref{bV3}) is
\begin{align}
bV_g(b,B)&\supset \int_0^\infty \d b' \int_0^{b'}\d\tau'\, {\sf T}(b\to b_k;b')V_{g}(b',B\setminus b_k)\\
&= \int_0^\infty b'\d b' \,{\sf T}(b\to b_k;b')V_{g}(b',B\setminus b_k)
\end{align}
where
\begin{align}
{\sf T}(b_1\to b_2;b_3) &\equiv \left(\text{\parbox{19em}{\centering\linespread{.5}\selectfont length of portion of bdy 1 s.t. $\gamma_p$ goes from bdy $1$ to bdy $2$ w/o leaving $\Lambda(b_1,b_2,b_3)$}}\right)\\ &= \log \frac{\cosh\frac{b_3}{2} + \cosh\frac{b_1+b_2}{2}}{\cosh\frac{b_3}{2}+\cosh\frac{b_1-b_2}{2}}.
\end{align}
Here we are using the notation $\Lambda(b_1,b_2,b_3)$ to mean a three-holed sphere with specified lengths. The explicit expression for ${\sf T}$ in the second line is the result of a simple calculation in the geometry of a hyperbolic three-holed sphere, see \cite{mirzakhani2007simple} or Appendix D.4 of \cite{Stanford:2019vob} for a derivation.

The sum of the second and third fates gives the same expression with $\sf{T}$ replaced by
\be
{\sf D}(b_1;b_2,b_3) \equiv \left(\text{\parbox{21em}{\centering\linespread{.5}\selectfont length of portion of bdy 1 s.t. $\gamma_p$ self-intersects or returns to bdy 1 w/o leaving $\Lambda(b_1,b_2,b_3)$}}\right).
\ee
The function ${\sf D}$ is determined by ${\sf T}$ and the sum rule
\be\label{sumrule}
b_1 = {\sf T}(b_1\to b_2;b_3) + {\sf T}(b_1\to b_3;b_2) + {\sf D}(b_1;b_2,b_3),
\ee
which expresses the fact that the possibilities represented by the terms on the RHS exhaust the possible fates of a geodesic on a three-holed sphere.

Combining these terms, and adding in the contributions from fates 4 and 5, we get Mirzakhani's recursion
\begin{align}
bV_g(b,B) =  &\sum_{k = 2}^{|B|}\int_0^\infty b'\d b' \,\left[b-{\sf T}(b\to b';b_k)\right]V_{g}(b',B\setminus b_k) \\
&+\frac{1}{2}\int_0^\infty b'\d b' b''\d b'' {\sf D}(b;b',b'') \left[V_{g-1}(b',b'',B) + \sum_{\text{stable}}V_{h_1}(b',B_1)V_{h_2}(b'',B_2)\right].
\end{align}
On the first line, we have combined fates 1, 2, and 3, using (\ref{sumrule}). On the second line, we have combined fates 4 and 5. The factor of one-half represents that we should not separately count the situation where we interchange the two gluing boundaries. The sum marked ``stable'' is over $h_1,h_2,B_1,B_2$ such that $h_1+h_2 = g$ and $B_1\cup B_2 = B$, omitting cases that would include $V_0(b')$ or $V_{0}(b',b_k)$.

As initial conditions for the recursion, one can use
\be\label{initialOrientable}
V_0(b_1,b_2,b_3) = 1,\hspace{20pt} V_{0}(b_1,b_2) = \frac{1}{b_1}\delta(b_1-b_2).
\ee
The second of these initial conditions is somewhat formal, but using it, one application of the recursion gives $V_{1}(b_1) = \frac{\pi^2}{12} + \frac{b_1^2}{48}$ which is a more conventional starting point for the recursion\footnote{The true volume of the moduli space with $(g,n) = (1,1)$ is actually $\pi^2/6 + b_1^2/24$. However, it is convenient to define $V_1(b_1)$ to be half this value. This is because the $(1,1)$ surface has a $\mathbb{Z}_2$ rotation symmetry, and with this definition, $b V_1(b)$ will be the true volume of the moduli space with a marked point on the boundary. The alternative to this is notationally awkward. Thanks to Zhenbin Yang for this explanation.} The reason we prefer the initial condition $V_0(b_1,b_2)$ is that it generalizes more straightforwardly to the non-orientable case.

\subsection{Non-orientable generalization}
In the non-orientable case, we need to modify the list of possible fates for geodesic $\gamma_p$. First of all, we leave in place all of the existing fates 1-5, with the additional condition that in each case, a neighborhood of $\gamma_p$ (together with the boundary that it begins on, which we will refer to as $b$) should be orientable. We also need to add in fates in which a neighborhood of $\gamma_p\cup b$ is non-orientable. In order for this neighborhood to be non-orientable, $\gamma_p$ must either self-intersect or return to the original boundary. Thickening the neighborhood of $\gamma_p\cup b$ will give a three-holed sphere $\tilde\Lambda$ with one hole given by $b$ and one hole closed off by a crosscap that $\gamma_p$ passes through. The third hole will be a gluing boundary (except in the exceptional case $\tilde\Lambda = X$ calculated in section \ref{special}). So the two possibilities are:
\begin{enumerate}
\item[6.] a neighborhood of $\gamma_p\cup b$ is non-orientable, $\gamma_p$ returns to $b$ within $\tilde{\Lambda}$
\be
\includegraphics[valign = c, scale = 1]{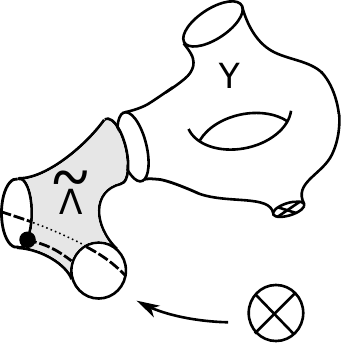}
\ee
\item[7.] a neighborhood of $\gamma_p$ is non-orientable, $\gamma_p$ self-intersects within $\tilde{\Lambda}$
\be
\includegraphics[valign = c, scale = 1]{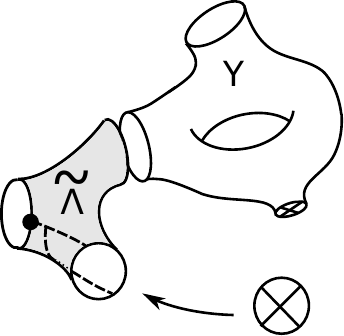}
\ee
\end{enumerate}

To include this fate, we need to know the analog of the ${\sf D}$ function for the non-orientable geometry $\tilde \Lambda$. Let $\tilde{\Lambda}(b_1,a,b_2)$ represent a three-holed sphere with hole sizes $b_1,a,b_2$, and with the hole of length $a$ closed off by a crosscap. Then we define
\be
{\sf C}(b_1,a;b_2) \equiv \left(\text{\parbox{21em}{\centering\linespread{1}\selectfont length of portion of bdy 1 s.t. a neighborhood of $\gamma_p$ is non-orientable and $\gamma_p$ self-intersects or returns to bdy 1 w/o leaving $\tilde{\Lambda}(b_1,a,b_2)$}}\right).
\ee
Integrating over $a$ with the crosscap measure, we get a function
\be\label{smallc}
{\sf c}(b_1;b_2) \equiv \int_0^\infty \frac{\d a}{2\tanh\frac{a}{4}}{\sf C}(b_1,a;b_2).
\ee
Deferring for a moment the computation of ${\sf C}$ and ${\sf c}$, the full recursion we will get after adding these fates in is 
\begin{align}
bV_g(b,B) =  &2\int_0^\infty b'\d b' \,\left[b-{\sf T}(b\to b';b_k)\right]V_{g}(b',B\setminus b_k) \\
&+\frac{1}{2}\int_0^\infty b'\d b' b''\d b'' {\sf D}(b;b',b'') \left[V_{g-1}(b',b'',B) + \sum_{\text{stable}}V_{h_1}(b',B_1)V_{h_2}(b'',B_2)\right]\label{suminline}\\
&+\frac{1}{2}\int_0^\infty b'\d b'{\sf c}(b;b')V_{g-\frac{1}{2}}(b',B).\label{finalTerm}
\end{align}
The first two lines are the same as in the orientable case, with the exception of a factor of two in the first line. The reason for this factor is that in the gluings of type 1,2,3, we can glue in a three-holed sphere that either does or does not have a reversal of orientation in going between the two new external boundaries. We also included a factor of $1/2$ in the last term to account for the $\mathbb{Z}_2$ mapping class group (MCG) of $\tilde{\Lambda}$.

To understand the MCG of $\tilde\Lambda$, and to evaluate ${\sf{C}}$ and ${\sf c}$, it is useful to consider the orientable double-cover of $\tilde\Lambda(b_1,a,b_2)$. We sketch it here with two different three-holed spheres shaded:
\be
\includegraphics[valign = c, scale = 1.75]{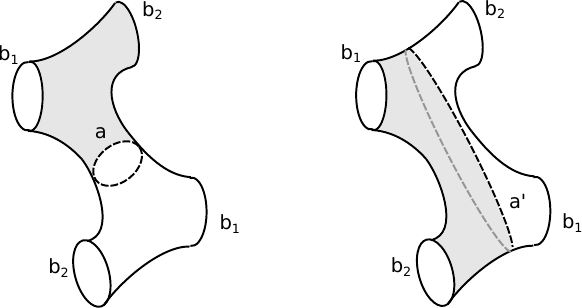}
\ee
Either of the shaded regions can be used as a fundamental region for the quotient, and the MCG interchanges the two. 

In order to evaluate ${\sf C}$, we are interested in the case where $\gamma_p$ starts on $b_1$ and a neighborhood of $\gamma_p\cup b_1$ is non-orientable. This is equivalent to requiring that $\gamma_p$ does not lie entirely in either of the two shaded regions of the orientable double cover. The length of the portion of the boundary such that $\gamma_p$ exits the shaded region of the left figure (while remaining in $\tilde\Lambda$) is ${\sf T}(b_1\to a;b_2)$. To further ensure that $\gamma_p$ neither makes it to $b_2$ nor remains in the shaded region of the right figure, we need to subtract off ${\sf T}(b_1\to b_2;a')$ and ${\sf D}(b_1;a',b_2)$, so
\begin{align}
{\sf C}(b_1,a;b_2) &= {\sf T}(b_1\to a;b_2) - {\sf T}(b_1\to b_2;a') - {\sf D}(b_1;a',b_2)\\
&= {\sf T}(b_1\to a;b_2) + {\sf T}(b_1\to a';b_2) -b_1.
\end{align}
In going to the second line, we used (\ref{sumrule}): note that the final expression is manifestly MCG-invariant under interchanging $a$ and $a'$. Here, $a'$ should be regarded as a function of $b_1,b_2,a$ due to the relationship between the lengths on the double cover \cite{Norbury}
\be
\cosh\frac{b_1}{2} + \cosh\frac{b_2}{2} = 2\sinh\frac{a}{4}\sinh\frac{a'}{4}.
\ee
With these definitions, the function ${\sf c}(b_1,b_2)$ is finite and can be computed explicitly
\begin{align}\label{c}
{\sf c}(b_1,b_2) &\equiv \int_0^\infty \frac{\d a}{2\tanh\frac{a}{4}}{\sf C}(b_1,a;b_2) \\
&= \int_0^{a_*}\frac{\d a}{\tanh\frac{a}{4}} {\sf T}(b_1\to a;b_2) + \int_{a_*}^{\infty}\frac{\d a}{\tanh\frac{a}{4}} ({\sf T}(b_1\to a;b_2)-b_1)\\
&= b_1b_2 - 2b_2 \log\frac{\cosh\frac{b_1+b_2}{4}}{\cosh\frac{b_1-b_2}{4}} + 4\left[\text{Li}_2(-e^{-\frac{b_1+b_2}{2}}) - \text{Li}_2(-e^{\frac{b_1-b_2}{2}})\right].\label{cans}
\end{align}
To get to the second line, we used invariance of the integral under $a\leftrightarrow a'$, and we defined $a_*$ as the value of $a$ such that $a = a'$. To verify the integral, one can first notice that the answer correctly vanishes for $b_1 = 0$. After taking the derivative with respect to $b_1$ (and remebering that this derivative also acts on $a_*$), the integral is straightforward in terms of the variable $s = \sinh\frac{a}{4}$.

This completes the derivation of the non-orientable recursion relation, but we still need to discuss the initial conditions. In genus zero, the non-orientable surfaces are the same as orientable ones, except that we are free to include orientation-reversing reflection operators at each boundary. So
\be
V_0^{\text{non-orientable}}(B) = 2^{|B|-1}V_0^{\text{orientable}}(B)
\ee
Here we have multiplied by two for each boundary component, and divided by two for the overall gauging of orientation reversal. We can use this, together with the initial conditions in the orientable case (\ref{initialOrientable}), to find the initial conditions
\begin{align}
V_0(b_1,b_2,b_3) = 4, \hspace{20pt} V_0(b_1,b_2) &= \frac{2}{b_1}\delta(b_1-b_2).
\end{align}

In the initial conditions we also need to account for the possibility in fates 4 or 5 that one of the components of $Y$ is a ``bare'' crosscap. This can be included by allowing the sum in line (\ref{suminline}) to include a formally-defined term $V_{1/2}(b')$, representing a bare crosscap. This should be defined so that when we glue to it with gluing measure $b_1\d b_1$, we get Norbury's measure for the crosscap. So
\begin{align}\label{v12initial}
V_{\frac{1}{2}}(b_1) &= \frac{1}{2b_1 \tanh\frac{b_1}{4}}.
\end{align}
Note that in this case, we should have the factor of $1/2$ written on line (\ref{suminline}), which accounts for the indistinguishability of the gluing boundaries $b',b''$, but we should not divide by a further $1/2$ to account for the MCG of the three-holed sphere with one hole closed off by a crosscap. The reason is that the geodesic $\gamma_p$ picks a particular MCG frame by avoiding one of the two possible choices of $a$ or $a'$. 

\subsection{Regularized volumes}
When the recursion is iterated, terms involving gluing to $V_{\frac{1}{2}}(b_1)$  lead to a logarithmic divergence for small $b_1$ --  the small-crosscap divergence described in the introduction. Gendulphe proposed \cite{Gendulphe} to avoid this divergence by restricting the moduli space so that all one-sided closed geodesics have length greater than $\epsilon/2$. For small $\epsilon$ we can compute these $\epsilon$-regularized volumes from the same recursion, but replacing (\ref{v12initial}) by
\be\label{modifiedIC}
V_{\frac{1}{2}}^{(\epsilon)}(b_1) = \frac{\theta(b_1>\epsilon)}{2b_1\tanh\frac{b_1}{4}}.
\ee
More precisely, for small $\epsilon$ a volume at genus $g$ will have the form
\be
V_{g}^{(\epsilon)}(b_1,\dots,b_n) = \sum_{k = 0}^{2g} \log(\epsilon^{-1})^k v_{g,k}(b_1,\dots,b_n) + O(\epsilon)
\ee
and the error between the volumes computed by our recursion (with initial condition (\ref{modifiedIC})) and Gendulphe's proposal will be contained in the $O(\epsilon)$ terms.\footnote{The reason there is any discrepancy to begin with (as pointed out to us by Norbury in an email) is that the recursion with these modified initial conditions will still receive some contributions from geometries with one sided geodesics smaller than $\epsilon/2$ (due to the possibility of forming a new short one-sided curve when gluing to $V_{g-1}$ in (\ref{suminline})). However, these contributions were integrable without the regulator, so the error we are making relative to Gendulphe's definition will vanish with $\epsilon$.}

In practice, it seems to be challenging to do the integrals and compute the $v_{g,k}$ functions, even for low genus. But as an example, we computed the volume of the genus-one moduli space with a geodesic boundary of zero length (cusp):
\be
V_{1}^{(\epsilon)}(0) = 2\log(\epsilon^{-1})^2 + 8\log(2)\log(\epsilon^{-1}) + \left[\frac{\pi^2}{4}+8\log(2)^2\right] + O(\epsilon).
\ee
For nonzero length the divergent terms are still fairly simple:
\begin{align}
V_{1}^{(\epsilon)}(b) &= 2  \log(\epsilon^{-1})^2 + \left\{\frac{b}{2} +4 \log(2 + 2e^{-b/2}) -  \frac{4}{b}\left[\frac{\pi^2}{6}+\text{Li}_2(-e^{-b/2}) + \text{Li}_2(-e^{b/2})\right]\right\}\log(\epsilon^{-1}) \notag \\ &\hspace{20pt}+ v_{1,0}(b) + O(\epsilon).
\end{align}
We did not compute $v_{1,0}(b)$ explictly, but it has the following limits
\be
v_{1,0}(0) = \frac{\pi^2}{4}+8\log(2)^2, \hspace{50pt} v_{1,0}(b\gg 1)\approx \frac{7 b^2}{48}
\ee
and it can be computed numerically:
\be
\includegraphics[valign = c, scale = .75]{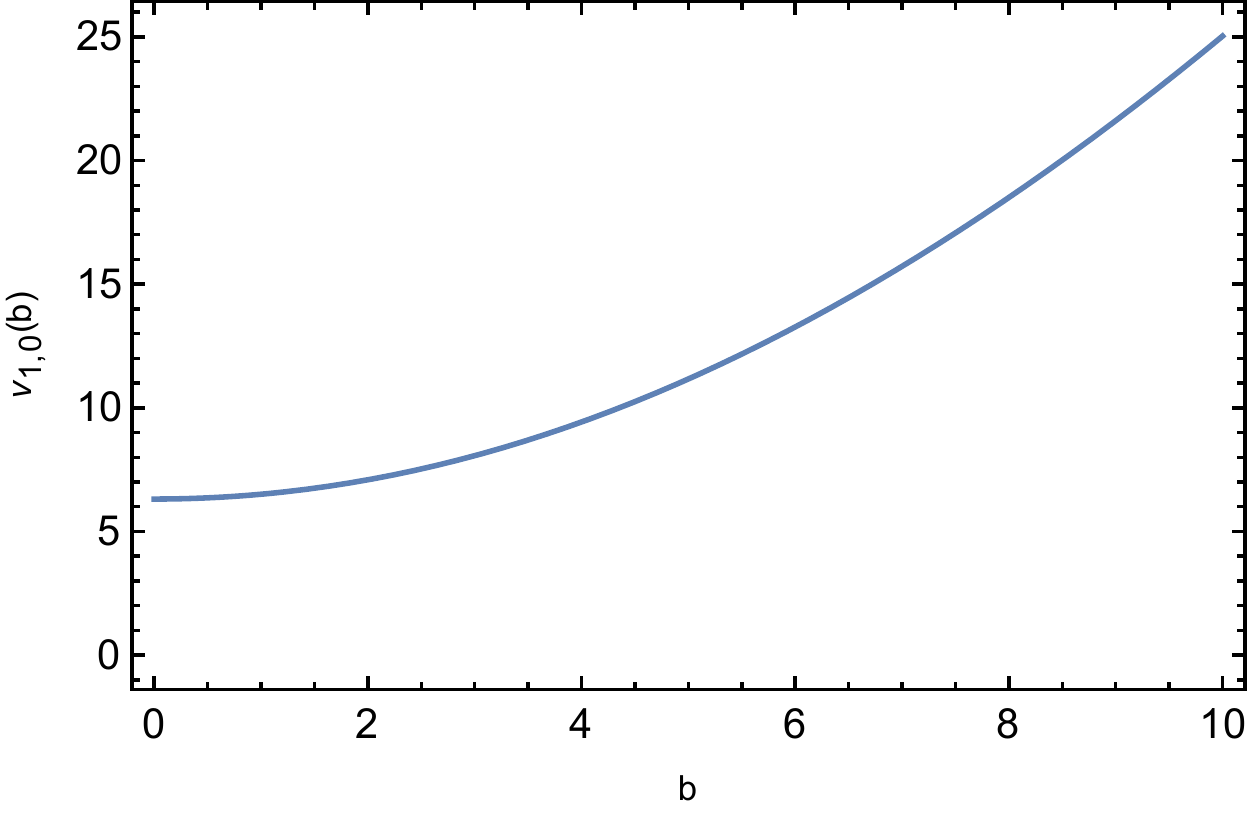}
\ee
Note that the $v_{g,k}(b_1,\dots,b_n)$ are sensitive to the precise regularization procedure. For example, if we required the lengths to be less than $\# \epsilon/2$ for some constant $\#\neq 1$, the $v_{g,k}(b_1,\dots,b_n)$ would transform by a triangular $k\times k$ matrix. Similarly, if we had chosen to expand the volumes in powers of e.g.~$\log(4/\epsilon)$ (this actually leads to slightly simpler expressions for the cases we computed), the $v_{g,k}$ would transform similarly. 

Also note that the leading power for large $b$ appears only in the $v_{1,0}$ term and is independent of $\epsilon$. In the orientable case, the coefficients of the leading large $b_i$ asymptotics of the moduli space volumes are intersection numbers, which can be computed by a matrix integral with spectral curve $y \sim z$.\footnote{This is referred to as the ``Airy'' or ``Kontsevich-Witten'' or ``topological'' model. For the connection to Weil-Petersson volumes see \cite{mirzakhani2007simple} or \cite{Dijkgraaf:2018vnm}. For the connection to the spectral curve $y\sim z$ see \cite{eynard2007invariants,Eynard:2014zxa} and for the connection to the low-energy limit of JT gravity see \cite{Saad:2022kfe}.} In the non-orientable case, a corollary of our result in section \ref{matrix} is that the large $b$ limit is computed by an orthogonal matrix integral with the same spectral curve $y \sim z$. This has a well-defined genus expansion without crosscap divergences. We don't know if there is an analog of the connection to intersection theory.

\subsection{Volume with two boundaries and one crosscap}\label{special}
In the way that we have organized the recursion, the case $V_{1/2}(b_1,b_2)$ is  exceptional, but it is easy enough to compute directly. The answer is just given by integrating over the size of the crosscap, where we fix the MCG by integrating $a$ only over the range of values where $a < a'$:
\be
V_{\frac{1}{2}}(b_1,b_2) = 2\int_0^{a_*}\frac{\d a}{2\tanh\frac{a}{4}}.
\ee
Here the factor of two is due the possbility of inserting a reflection operator, and
\be\label{astar}
2\sinh^2\frac{a_*}{4} = \cosh\frac{b_1}{2} + \cosh\frac{b_2}{2}.
\ee
There is a small-crosscap divergence at $a = 0$, and the $\epsilon$-regularized answer is
\begin{align}
V_{\frac{1}{2}}^{(\epsilon)}(b_1,b_2)&= 2\int_\epsilon^{a_*}\frac{\d a}{2\tanh\frac{a}{4}}\\
&= 2\log\left[\frac{\cosh\frac{b_1}{2} + \cosh\frac{b_2}{2}}{2}\right] - 4\log\left[\sinh\frac{\epsilon}{4}\right]\\
&\approx 2\log\left[\frac{\cosh\frac{b_1}{2} + \cosh\frac{b_2}{2}}{2}\right] - 4\log\left[\frac{\epsilon}{4}\right].
\end{align}
See section 3.2 of \cite{Saad:2022kfe} for some discussion of this formula.

\section{Matching to a matrix integral recursion relation}\label{matrix}
Mirzakhani's recursion implies that JT gravity is dual to a matrix integral. A key fact that underlies this statement was Eynard and Orantin's discovery \cite{eynard2007weil} that Laplace transforms of the volumes $V_g(B)$ satisfy the same type of recursion relation as the loop equations of a matrix integral with unitary symmetry class. This type of recursion is known as topological recursion \cite{eynard2007invariants}, and it is characterized by a hyperelliptic spectral curve
\be
x = -z^2, \hspace{20pt} y = \frac{\sin(2\pi z)}{4\pi}.
\ee
In matrix integral language, specifying the spectral curve amounts to specifying the leading density of eigenvalues: $\rho(x) \propto \i y(x) \propto \sinh(2\pi \sqrt{x})$.

In \cite{Stanford:2019vob}, it was conjectured that JT gravity with non-orientable surfaces included should be related to a matrix integral with the same spectral curve, but with symmetry class corresponding to conjugation by orthogonal matrices. For this symmetry class, the loop equations still make sense, but they do not reduce to topological recursion. Mathematically, the conjecture is that the recursion implied by the loop equations agrees with the recursion satisfied by the Laplace transforms of the volumes of non-orientable moduli space.

We will now show this is true. We define the Laplace transforms of the volumes as ``resolvents'' $R_g(x_1,\dots,x_n) = R_g(-z_1^2,\dots, -z_n^2)$
\be
R_g(-z_1^2,\dots,-z_n^2) = (-1)^n \int_0^\infty V_g(b_1,\dots,b_n)\prod_{j = 1}^n b_j\d b_j \frac{e^{-b_j z_j}}{2z_j}.
\ee
The inverse relationship is
\be
V_g(b_1,\dots,b_n) = (-1)^n \int_{\delta + \i \mathbb{R}}R_g(-z_1^2,\dots,z_n^2)\prod_{j = 1}^n \frac{\d z_j}{2\pi \i}\frac{2 z_j}{b_j} e^{b_j z_j}.
\ee

The loop equations of a matrix integral with orthogonal symmetry class are the case $\upbeta = 1$ in section 4.1 of \cite{Stanford:2019vob}. The recursion relation is
\be
R_g(-z^2,I) = \frac{1}{2\pi \i z} \int_{\delta - \i \infty}^{\delta + \i \infty}\frac{z'^2\d z'}{z'^2-z^2}\frac{F_g(-z'^2,I)}{y(-z'^2)},
\ee
where $I$ represents $\{x_1,\dots,x_n\}$ or equivalently  $\{-z_1^2,\dots,-z_n^2\}$ and 
\begin{align}\label{Fg}
F_g(x,I) \equiv &-\partial_x R_{g-\frac{1}{2}}(x,I) + R_{g-1}(x,x,I) + \sum_{\text{stable}}R_h(x,J)R_{g-h}(x,I\setminus J)\\
&+ \sum_{k = 1}^{|I|}\left(\frac{1}{2\sqrt{-x}\sqrt{-x_k}(\sqrt{-x}+\sqrt{-x_k})^2} + \frac{1}{(x-x_k)^2}\right)R_g(x,I\setminus x_k).
\end{align}
Our goal is to start from this equation and derive the non-orientable Mirzakhani recursion. We can save some work by comparing to the analogous statement in the orientable case. There the recursion is similar except the first term (involving $-\partial_x R_{g-\frac{1}{2}}(x,I)$) is absent, and the last term has an additional factor of $\frac{1}{2}$. This recursion is known \cite{eynard2007weil} to agree with Mirzakhani's recursion in the orientable case. Our non-orientable recursion has all of the terms present in the orientable recursion, and an extra factor of two in the term involving a sum over $k$, so these terms will match.

So all that remains is to show that the first term in (\ref{Fg}) maps onto the final term (\ref{finalTerm}). This term reads
\be
R_g(-z^2,I) \supset \frac{1}{2\pi \i z}\int_{\delta + \i \mathbb{R}} \frac{z'^2\d z'}{z'^2-z^2}\frac{1}{y(-z'^2)}\frac{1}{2z'}\partial_{z'}R_{g-\frac{1}{2}}(-z'^2,I).
\ee
Inserting the Laplace and inverse Laplace transforms, we find
\begin{align}
V_{g}(b,B) &\supset \int_{2\delta + \i \mathbb{R}}\frac{\d z}{2\pi \i}\frac{2 z}{b}e^{b z}\frac{1}{2\pi \i z}\int_{\delta + \i \mathbb{R}}\frac{z'^2\d z'}{z'^2-z^2}\frac{1}{y(-z'^2)}\frac{1}{2z'}\partial_{z'}\int_0^\infty b'\d b'V_{g-\frac{1}{2}}(b',B)\frac{e^{-b'z'}}{2 z'}.
\end{align}
The $z$ integral can be done by contour integration, picking up residues from the poles at $z = \pm z'$. The $z'$ and $b'$ integrals can then be interchanged, and we find (after multiplying through by $b$)
\begin{align}
bV_g(b,B) &=\frac{1}{2}\int_0^\infty b'\d b'V_{g-\frac{1}{2}}(b',B)\underbrace{\int_{\delta + \i \mathbb{R}}\frac{2\d z'}{\i}\frac{\sinh(b z')}{\sin(2\pi z')}\left(-\partial_{z'}\frac{e^{-b'z'}}{ z'}\right)}_{\text{this equals }{\sf c}(b,b')\text{ in }(\ref{cans})}.
\end{align}
The agreement between this recursion and the non-orientable Mirzkhani recursion boils down to the fact that the underlined expression is equal to the ${\sf c}(b,b')$ defined in (\ref{c}). This integral can be evaluated using
\be\label{id}
\int_{\delta + \i \mathbb{R}}\frac{\d z'}{\i}\frac{e^{-b z'}}{\sin(2\pi z')} = \frac{1}{e^{b/2}+1}
\ee
together with two integrals of this expression with respect to $b$. The identity (\ref{id}) is a straightforward application of the residue theorem

\section*{Acknowledgments}
We are grateful to Adel Rahman for initial collaboration and to Paul Norbury for comments. This research was supported in part by DOE grant DE-SC0021085, by the Sloan Foundation, by a grant from the Simons foundation (926198, DS), and by the National Science Foundation under Grant No.~NSF PHY-1748958.

\bibliography{references}

\providecommand{\href}[2]{#2}\begingroup\raggedright\begin{thebibliography}{10}

\bibitem{mirzakhani2007simple}
M.~Mirzakhani, ``Simple geodesics and {Weil-Petersson} volumes of moduli spaces
  of bordered riemann surfaces,'' {\em Inventiones mathematicae} {\bfseries
  167} no.~1, (2007) 179--222.

\bibitem{Norbury}
P.~Norbury, ``Lengths of geodesics on non-orientable hyperbolic surfaces,''
  {\em Geometriae Dedicata} {\bfseries 134} no.~1, (2008) 153--176.

\bibitem{Gendulphe}
M.~Gendulphe, ``What's wrong with the growth of simple closed geodesics on
  nonorientable hyperbolic surfaces,'' {\em arXiv preprint arXiv:1706.08798}
  (2017) .

\bibitem{Stanford:2019vob}
D.~Stanford and E.~Witten, ``{JT gravity and the ensembles of random matrix
  theory},'' \href{http://dx.doi.org/10.4310/ATMP.2020.v24.n6.a4}{{\em Adv.
  Theor. Math. Phys.} {\bfseries 24} no.~6, (2020) 1475--1680},
  \href{http://arxiv.org/abs/1907.03363}{{\ttfamily arXiv:1907.03363
  [hep-th]}}.

\bibitem{eynard2007weil}
B.~Eynard and N.~Orantin, ``Weil-{P}etersson volume of moduli spaces,
  {M}irzakhani's recursion and matrix models,''
  \href{http://arxiv.org/abs/0705.3600}{{\ttfamily arXiv:0705.3600 [math-ph]}}.

\bibitem{Saad:2019lba}
P.~Saad, S.~H. Shenker, and D.~Stanford, ``{JT gravity as a matrix integral},''
\href{http://arxiv.org/abs/1903.11115}{{\ttfamily arXiv:1903.11115 [hep-th]}}.

\bibitem{Teitelboim:1983ux}
C.~Teitelboim, ``{Gravitation and Hamiltonian Structure in Two Space-Time
  Dimensions},''
\href{http://dx.doi.org/10.1016/0370-2693(83)90012-6}{{\em Phys. Lett.}
  {\bfseries B126} (1983) 41--45}.

\bibitem{Jackiw:1984je}
R.~Jackiw, ``{Lower Dimensional Gravity},''
\href{http://dx.doi.org/10.1016/0550-3213(85)90448-1}{{\em Nucl. Phys.}
  {\bfseries B252} (1985) 343--356}.

\bibitem{Dijkgraaf:2018vnm}
R.~Dijkgraaf and E.~Witten, ``{Developments in Topological Gravity},''
  \href{http://dx.doi.org/10.1142/S0217751X18300296}{{\em Int. J. Mod. Phys. A}
  {\bfseries 33} no.~30, (2018) 1830029},
  \href{http://arxiv.org/abs/1804.03275}{{\ttfamily arXiv:1804.03275
  [hep-th]}}.

\bibitem{eynard2007invariants}
B.~Eynard and N.~Orantin, ``Invariants of algebraic curves and topological
  expansion,'' \href{http://arxiv.org/abs/math-ph/0702045}{{\ttfamily
  arXiv:math-ph/0702045}}.

\bibitem{Eynard:2014zxa}
B.~Eynard, ``{A short overview of the "Topological recursion"},''
\href{http://arxiv.org/abs/1412.3286}{{\ttfamily arXiv:1412.3286 [math-ph]}}.

\bibitem{Saad:2022kfe}
P.~Saad, D.~Stanford, Z.~Yang, and S.~Yao, ``{A convergent genus expansion for
  the plateau},'' \href{http://arxiv.org/abs/2210.11565}{{\ttfamily
  arXiv:2210.11565 [hep-th]}}.

\end{thebibliography}\endgroup

\bibliographystyle{utphys}

\end{document}